\documentclass[aps, prb, twocolumn,superscriptaddress,preprintnumbers,amsmath,amssymb]{revtex4-2}
\usepackage{graphicx}
\usepackage{dcolumn}
\usepackage{bm}
\usepackage{siunitx}
\usepackage{float}
\usepackage{graphicx,here}

\usepackage{multirow}
\usepackage{color}
\usepackage{soul}

\usepackage{ulem}
\usepackage{xcolor}
\DeclareRobustCommand{\erase}{\bgroup\markoverwith{\textcolor{red}{\rule[.5ex]{2pt}{0.4pt}}}\ULon}

\begin{document}

\title{Structural 130-K Phase Transition and Emergence of a Two-Ion Kondo State in 
HT-Ce$_2$Rh$_2$Ga Explored by $^{69,71}$Ga Nuclear Quadrupole Resonance}

\author{Sh. Yamamoto}
\email{s.yamamoto@hzdr.de}
\affiliation{Hochfeld-Magnetlabor Dresden (HLD-EMFL) and W\"{u}rzburg-Dresden Cluster of Excellence ct.qmat, Helmholtz-Zentrum Dresden-Rossendorf, 01328 Dresden, Germany}
\affiliation{Max Planck Institute for Chemical Physics of Solid, D-01187 Dresden, Germany}
\author{T. Fujii}
\affiliation{Max Planck Institute for Chemical Physics of Solid, D-01187 Dresden, Germany}
\author{S. Luther}
\affiliation{Hochfeld-Magnetlabor Dresden (HLD-EMFL) and W\"{u}rzburg-Dresden Cluster of Excellence ct.qmat, Helmholtz-Zentrum Dresden-Rossendorf, 01328 Dresden, Germany}
\affiliation{Institut f$\ddot{u}$r Festk$\ddot{o}$rper- und Materialphysik, TU Dresden, 01062 Dresden, Germany}
\author{H. Yasuoka}
\affiliation{Max Planck Institute for Chemical Physics of Solid, D-01187 Dresden, Germany}
\author{H. Sakai}
\affiliation{Advanced Science Research Center, Japan Atomic Energy Agency, Tokai, Ibaraki 319-1195, Japan}
\author{F. B\"{a}rtl}
\affiliation{Hochfeld-Magnetlabor Dresden (HLD-EMFL) and W\"{u}rzburg-Dresden Cluster of Excellence ct.qmat, Helmholtz-Zentrum Dresden-Rossendorf, 01328 Dresden, Germany}
\affiliation{Institut f$\ddot{u}$r Festk$\ddot{o}$rper- und Materialphysik, TU Dresden, 01062 Dresden, Germany}
\author{K. M. Ranjith}
\affiliation{Max Planck Institute for Chemical Physics of Solid, D-01187 Dresden, Germany}
\author{H. Rosner}
\affiliation{Max Planck Institute for Chemical Physics of Solid, D-01187 Dresden, Germany}
\author{J. Wosnitza}
\affiliation{Hochfeld-Magnetlabor Dresden (HLD-EMFL) and W\"{u}rzburg-Dresden Cluster of Excellence ct.qmat, Helmholtz-Zentrum Dresden-Rossendorf, 01328 Dresden, Germany}
\affiliation{Institut f$\ddot{u}$r Festk$\ddot{o}$rper- und Materialphysik, TU Dresden, 01062 Dresden, Germany}
\author{A. M. Strydom}
\affiliation{Max Planck Institute for Chemical Physics of Solid, D-01187 Dresden, Germany}
\affiliation{Highly Correlated Matter Research Group, Physics Department, University of Johannesburg, PO Box 524, Auckland Park 2006, South Africa}
\author{H. K\"{u}hne}
\affiliation{Hochfeld-Magnetlabor Dresden (HLD-EMFL) and W\"{u}rzburg-Dresden Cluster of Excellence ct.qmat, Helmholtz-Zentrum Dresden-Rossendorf, 01328 Dresden, Germany}
\author{M. Baenitz}
\affiliation{Max Planck Institute for Chemical Physics of Solid, D-01187 Dresden, Germany}

\date{\today}

\begin{abstract}

We have studied the microscopic magnetic properties, the nature of the 130-K phase transition, and the ground state in the recently synthesized compound Ce$_2$Rh$_2$Ga by use of $^{69,71}$Ga nuclear quadrupole resonance (NQR). The NQR spectra clearly show an unusual phase transition at $T_t$ $\sim$ 130 K yielding a splitting of the high-temperature single NQR line into two clearly resolved NQR lines, providing evidence for two crystallographically inequivalent Ga sites. The NQR frequencies are in good agreement with fully-relativistic calculations of the band structure. Our NQR results indicate the absence of magnetic or charge order down to 0.3 K. The temperature dependence of the spin-lattice relaxation rate, 1/$T_1$, shows three distinct regimes, with onset temperatures at $T_t$ and 2 K. The temperature-independent 1/$T_1$, observed between $T_t$ and 2 K, crosses over to a Korringa process, 1/$T_1$ $\propto$ $T$, below $\sim$ 2 K, which evidences a rare two-ion Kondo scenario: the system goes into a dense Kondo coherent state at 2.0 and 0.8 K for the two different Ga sites.

\end{abstract}

\maketitle


\textit{Introduction.} --- Among the correlated 4$f$-ion systems, there are only a few compounds in which a structural instability at high temperatures has an effect on the ground state and especially on the correlations among the 4$f$ ions and conduction electrons. While in Yb systems charge ordering or even significant intermediate-valence behavior can occur, Ce systems are usually characterized by a rather stable valence \mbox{\cite{brandt1984concentrated,hewson1985models}}. Ce$_2$Rh$_2$Ga seems to be an exception to the rule. This material is dimorphic, with high- (HT) and low-temperature (LT) forms, yielding an orthorhombic La$_2$Ni$_3$-type (space group $Cmce$) and monoclinic (space group $C2/$$c$) structure at room temperature, respectively, depending on thermal treatments during the sample synthesis \cite{nesterenko2020two}. 

In the so-called HT-Ce$_2$Rh$_2$Ga form a structural phase transition at 130 K changes the ground state due to a reconstruction of the Fermi surface and a slight increase of the Ce valence beyond 3+ \mbox{\cite{nesterenko2020two,dudka2022multi,sato2022valence}}. The question arises how this influences the coupling of charge and spin. As a consequence of the 130-K phase transition, the coupling between the cerium ions is reduced \cite{YamamotoSupplemental}, which is a prerequisite for the formation of the heavy-fermion state. Furthermore, the low-temperature phase exhibits two inequivalent Ce ions, which are exposed to Kondo screening at different temperatures. Multi-ion Kondo physics is rare and only poorly explored. Exceptions are theoretical studies \mbox{\cite{lee1986theories,benlagra2011kondo,jiang2020enhanced}} as well as experimental studies of site-dependent magnetic transitions for Ce$_3$Pd$_{20}$Si$_6$ \mbox{\cite{paschen2007quantum}}, Ce$_3$PtIn$_{11}$ \mbox{\cite{proklevska2015magnetism,kambe2020115,fukazawa2020successive}}, and Ce$_{7}$Ni$_{3}$ \mbox{\cite{umeo2003field}}. Here, site-selective microscopic experiments can provide crucial information for understanding the underlying physics.  

In this Letter, we study the HT form, HT-Ce$_2$Rh$_2$Ga, which exhibits a unique phase transition at $T_t$ $\sim$ 130 K that is accompanied by signatures in the specific heat and susceptibility. Taking these anomalies into account, the authors in Ref. \cite{nesterenko2020two} argued that below $T_t$, HT-Ce$_2$Rh$_2$Ga undergoes an antiferromagnetic (AFM) order. Alternatively, a charge density wave (CDW) scenario is also possible. However, the origin of the 130-K phase transition is not yet fully understood. 

The structural phase change across $T_t$ has been studied by x-ray diffraction experiments \cite{dudka2022multi}. The structure changes from a room-temperature orthorhombic phase ($a$ = 5.845 \AA, $b$ = 9.573 \AA, $c$ = 7.496 \AA) to a monoclinic phase with non-merohedral twinning of the space group $C2/m$ ($a$ = 9.401 \AA, $b$ = 5.807 \AA, $c$ = 7.595 \AA, $\beta$ = 91.89$^\circ$). In the monoclinic phase below $T_t$, two inequivalent Ga sites are expected \cite{dudka2022multi,YamamotoSupplemental}. Recent resonant x-ray emission spectroscopy experiments indicate that the average Ce valence, $v$, increases by $\sim$ 0.7 $\%$ from the orthorhombic ($v$ $\sim$ 3.053) to the monoclinic ($v$ $\sim$ 3.075) phase \cite{sato2022valence}. In addition, our specific-heat data indicate the occurrence of a heavy-fermion state \mbox{\cite{YamamotoSupplemental}}. 

The aim of this Letter is to provide microscopic information on the nature of the 130-K phase transition, as well as on the formation of the correlated ground state. The present NQR results clearly show that neither an AFM nor CDW scenario can be considered. The symmetry reduction below 130 K yields two magnetically inequivalent Ce and Ga sites in the lattice. The NQR spin-lattice relaxation measured at these two Ga sites clearly indicates the formation of a heavy-fermion state, but the transition to this occurs at two distinct temperatures. Thus, in CeRh$_2$Ga$_2$ there exist two cerium ions which are differently shielded by the surrounding conduction electrons (two-ion Kondo effect). The two cerium sublattices are magnetically coupled to each other. HT-Ce$_2$Rh$_2$Ga is, thus, a rather unusual 4$f$ system which can serve as a platform to study the two-ion Kondo physics.




\begin{figure}[t]
	\begin{center}
		\includegraphics[width=0.99\columnwidth]{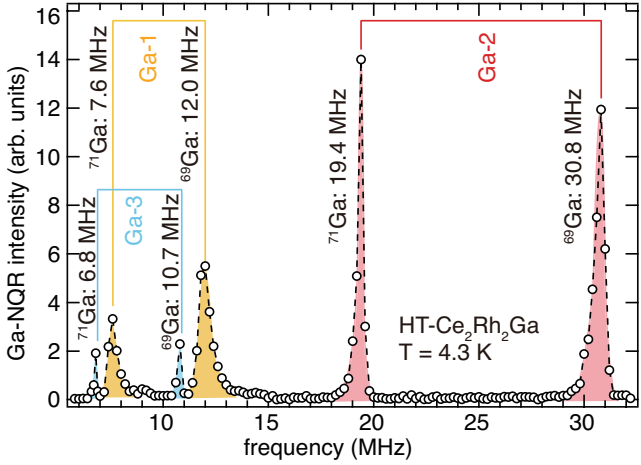}
	\end{center}
	\caption{Ga-NQR spectra of HT-Ce$_2$Rh$_2$Ga at 4.3 K. There are three pairs of $^{69}$Ga and $^{71}$Ga NQR signals, Ga-1, 2, and 3, that are assigned to three inequivalent Ga sites.}
	\label{GaWideSpectrum}
\end{figure}

\textit{Experimental.} --- In this study, we focused on the Ga NQR in HT-Ce$_2$Rh$_2$Ga, where each Ce atom is surrounded by four Ga neighbors, forming edge-sharing CeGa$_4$ tetrahedra \cite{YamamotoSupplemental}. The sample was prepared in the same way as described in Ref. \cite{nesterenko2020two}, i.e., the arc-melted ingot was annealed at 900 $^\circ$C for 30 days in an evacuated quartz tube and then quenched in cold water. The annealed ingot was checked by x-ray diffraction to ensure the HT-Ce$_2$Rh$_2$Ga existence of the phase, and then powdered to about 200 $\mu$m particle size. The powdered sample was mixed with paraffin so that the skin-depth effect due to eddy currents by radio-frequency excitation during the NQR measurements is negligible \cite{YamamotoSupplemental}. 

The Ga nuclei have two stable isotopes, $^{69}$Ga and $^{71}$Ga. Both of them have a nuclear spin of $I$ = 3/2, and the nuclear gyromagnetic ratio, $\gamma$, the nuclear quadrupole moment, Q, and the natural abundance, $A$, are $^{69}\gamma$ = 10.2192 MHz/T, $^{69}$Q = 0.178 barns, $^{69}A$ = 60 \% for $^{69}$Ga, and $^{71}\gamma$ = 12.9847 MHz/T, $^{71}$Q = 0.112 barns, $^{71}A$ = 40 \%  for $^{69}$Ga, respectively. The ratios of $\gamma$ and Q between the two isotopes are ($^{69}\gamma$/$^{71}\gamma$) = 0.7870 and ($^{69}$Q/$^{71}$Q) = 1.589. The NQR spectra and the nuclear relaxation times were measured using a standard pulsed (spin-echo) NMR apparatus (TecMag-Apollo). The NQR spectra were recorded by the frequency-sweep method. Further experimental details are described in Note 2 of the Supplemental Material \cite{YamamotoSupplemental}. 

\textit{Ga NQR spectra.} --- For searching the NQR signal, we consider the nuclear energy levels, $E_m$, obtained from the nuclear quadrupole Hamiltonian which can be expressed in the case of a nonaxial field gradient for $I$ = 3/2 as \cite{DasNuclear1958},

\begin{eqnarray}
E_{\pm3/2} &=& \frac{1}{2}h\nu_\mathrm{Q}\sqrt{1+\frac{\eta^2}{3}},\\
E_{\pm1/2} &=& -\frac{1}{2}h\nu_\mathrm{Q}\sqrt{1+\frac{\eta^2}{3}},\\
\nu_\mathrm{Q} &\equiv& \frac{3e^2q\mathrm{Q}}{h2I(2I-1)},
\label{etaZeroEnergy}
\end{eqnarray}

\begin{figure}[t]
	\begin{center}
		\includegraphics[width=0.99\columnwidth]{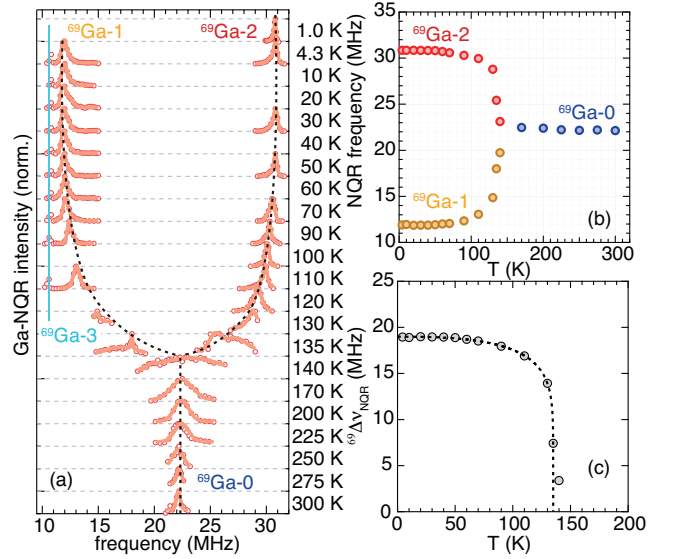}
	\end{center}
	\caption{(a)Temperature dependence of the $^{69}$Ga-NQR spectra of HT-Ce$_2$Rh$_2$Ga. $^{69}$Ga-0 is associated with the single Ga site in the orthorhombic phase ($T$ $>$ $T_t$), and the $^{69}$Ga-1 and $^{69}$Ga-2 signals stem from two inequivalent Ga sites in the monoclinic phase ($T$ $<$ $T_t$). (b) Temperature dependence of the $^{69}$Ga-NQR peak frequencies extracted by using a Gaussian fit. Blue circles are data from $^{69}$Ga-0 above $T_t$. Yellow and red circles are data from $^{69}$Ga-1 and $^{69}$Ga-2, respectively. (c) Temperature dependence of $^{69}\Delta\nu_\mathrm{NQR}$ = $f_\mathrm{Q}$(Ga-2)-$f_\mathrm{Q}$(Ga-1). The dashed line is a guide to the eye.}
	\label{GaNQRTempDep}
\end{figure}

\begin{table*}[t]
\caption{Calculated EFG tensors and quadrupole coupling constants in HT-Ce$_2$Rh$_2$Ga (see main text). Ga-0 is associated with the Ga site in the orthorhombic phase above $T_t$. Ga-1 and Ga-2 are two inequivalent Ga sites in the monoclinic phase below $T_t$, respectively. The Wyckoff position for each Ga site is also shown.}
	\begin{ruledtabular}
		\begin{tabular}{lccccc}
			\multirow{2}{*}{Ga sites} &\multicolumn{4}{c}{Calculation} & Experiment\\
			\cline{2-5}
			\cline{6-6}
			 & $V_{ZZ} = \partial V/\partial x_Z\partial x_Z$ (10$^{21}$ V/m$^2$) & $\eta$ & $\nu_\mathrm{Q}$ (MHz) & $\nu_\mathrm{NQR}$ (MHz) & $\nu_\mathrm{NQR}$ (MHz)\\
			\hline
			Ga-0 ($Cmce$, 4$a$)  & 10.199 & 0.790 &22.693 &24.124 & 22.336\\
			Ga-1 ($C2/m$,2$a$)  & -6.609 & 0.854 &14.705 & 15.858 &11.964\\
			Ga-2 ($C2/m$,2$d$)  & 13.197 & 0.652 &29.363 & 30.346 &30.758\\
		\end{tabular}
	\end{ruledtabular}
	\label{Table:EFGCal}
\end{table*}

\noindent where $\nu_{\mathrm{Q}}$ denotes the quadrupole coupling between the electric-field gradient (EFG) stemming from an asymmetric charge distribution at the nuclear site, $q$, and  the nuclear quadrupole moment, Q. The NQR occurs for the transition between two levels $m$ and $m$+1, and the NQR frequency can be written as, $f_\mathrm{Q}$ = $\nu_\mathrm{Q}$(2$m$+1)/2. Hence, only one NQR line for $I$ = 3/2, $^{69}$Ga and $^{71}$Ga, is expected. 

In Fig. \ref{GaWideSpectrum}, we show a broad-band Ga NQR spectrum at 4.3 K (in the monoclinic phase). We observe three pairs of $^{69}$Ga and $^{71}$Ga signals labeled Ga-1, Ga-2, and Ga-3. The observed frequency ratios for the NQR pairs $^{69}$Ga and $^{71}$Ga NQR are 1.572 $\pm$ 0.005, 1.589 $\pm$ 0.002 and 1.588 $\pm$ 0.010, respectively. These numbers agree very well with the ratio of $^{69}$Q/$^{71}$Q = 1.589, confirming the validity of the spectral assignment. 

Focusing on the $^{69}$Ga NQR signal, we show the temperature dependence of the NQR spectrum in Fig. \ref{GaNQRTempDep}(a), including data above $T_t$ (in the orthorhombic phase). The single Ga-NQR signal, $^{69}$Ga-0, shows spectral broadening with decreasing temperature from 300 to 170 K, presumably due to some inhomegeneous distribution of the phase transition. Below $\sim$ 130 K, the single peak Ga-0 splits into two lines, Ga-1 and Ga-2. We observe an additional Ga-NQR line, $^{69}$Ga-3, at 10.8 MHz. Since the $^{69}$Ga-3 signal yields a constant temperature dependence, we associate this with a signal either from twin boundaries or an impurity phase. The Ga-1 (Ga-2) peak shifts to lower (higher) frequency with decreasing temperature, reaching 12 MHz (30.8 MHz) at 4.3 K, see Fig. \ref{GaNQRTempDep}(b). 

\begin{figure}[t]
	\begin{center}
		\includegraphics[width=0.95\columnwidth]{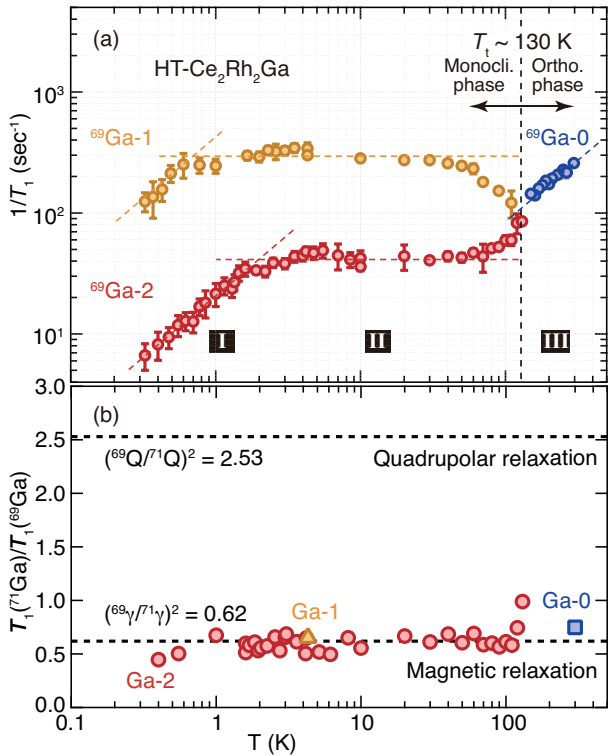}
	\end{center}
	\caption{(a) Temperature dependence of the $^{69}$Ga-NQR spin-lattice relaxation rate, 1/$T_1$, for the $^{69}$Ga-0 (blue), $^{69}$Ga-1 (yellow) and $^{69}$Ga-2 (red) lines. There are three distinct temperature regions where the relaxation processes yield a different characteristic behavior. Region III is in the orthorhombic phase above $T_t$ $\sim$ 130 K and 1/$T_1$ behaves like for a Korringa process. Region I and II lie within the monoclinic phase below $T_t$. (b) The ratio of $T_1$ for the two Ga isotopes, $^{71}$Ga/$^{69}$Ga, for the Ga-0, Ga-1, and Ga-2 lines are shown by blue, yellow, and red symbols, respectively.}
	\label{TOneRateTempDep}
\end{figure}

A question arises concerning the origin of the line splitting below $T_t$, because the first report of Ce$_2$Rh$_2$Ga proposed an AFM order below $T_t$ \cite{nesterenko2020two}. However, if this would be the case, we would see an NQR line splitting due to the local hyperfine field from ordered Ce moments. As clearly seen in Fig. \ref{GaNQRTempDep}, we observe very sharp NQR lines for each Ga-1 and Ga-2 site at low temperatures, definitively affirming the absence of AF order. Further, as shown in the following, the temperature dependence of the spin-lattice relaxation rate does not display any peaks related to magnetic order [Fig. \ref{TOneRateTempDep}(a)], which excludes a possible scenario in which local fields from the Ce ions cancel each other at the Ga site. Preliminary $\mu$SR measurements also support the absence of magnetic order above 2 K \mbox{\cite{strydomprivate}}. Furthermore, any spin and/or charge density wave order is excluded. In these cases, we should see a characteristic inhomogeneous line broadening due to the EFG or internal field modulation. 

In order to check the site assignment above we performed EFG calculations based on the band structure calculated using the density functional theory solid-state code \textsc{fplo} \mbox{\cite{koepernik1999full}}. The $k$-meshes in both structures were carefully converged with respect to the total energy and the calculated EFG. We used the Perdew-Wang parametrization of the local density approximation for the exchange-correlation functional \mbox{\cite{perdew1992accurate}}. To account for the strong electronic correlations of the Ce 4$f$ electrons, we applied the LDA+U approach with U = 6 eV for the Ce 4$f$ states. Varying U by $\pm$ 1 eV did not show significant changes. Alternatively, treating the Ce 4$f$ electron as a core state in the frozen core approximation yields very similar results for the EFG's on the Ga sites. The quadrupole coupling, $\nu_\mathrm{Q}$, can be obtained by calculating the EFG at each Ga nuclear site which is defined as the second partial derivative of the electrostatic potential at the position of the nucleus. We summarized the obtained maximum EFG, $V_{ZZ}$, asymmetry parameter [$\eta$  = ($V_{XX}$-$V_{YY}$)/$V_{ZZ}$], and the quadrupole-coupling constant $\nu_\mathrm{Q}$ in Table \mbox{\ref{Table:EFGCal}}. We compare then the extracted NQR frequencies with the experimental values. Thereby, we used $\nu_\mathrm{NQR}$ = $\nu_\mathrm{Q}$(1+$\eta^2$/3)$^{1/2}$ for $I$ = 3/2. The NQR frequencies obtained from our experiments are in good agreement with the calculated values, assuring that the site assignment is correct.

Our specific-heat and susceptibility results signal a first-order transition in a narrow temperature region around $T_t$ \cite{nesterenko2020two,YamamotoSupplemental}. On the other hand, a gradual variation of the NQR frequency has been observed from $T_t$ down to $\sim$ 80 K. Below $T_t$, the variation of the NQR frequencies couples predominantly to the evolving structural monoclinicity. The temperature-dependent splitting of the NQR spectra can be evaluated as $^{69}\Delta\nu_\mathrm{NQR}$ = $f_\mathrm{Q}$(Ga-2)-$f_\mathrm{Q}$(Ga-1), and is depicted in Fig. \ref{GaNQRTempDep}(c). The lattice parameters \cite{nesterenko2020two}, and Ce 4$f$ valence \cite{sato2022valence} also display a continuous variation with respect to temperature similar to that of the NQR frequency below $T_t$.

\textit{Nuclear magnetic relaxation.} --- The nuclear spin-lattice relaxation time, $T_1$, has been measured by the inversion-recovery method, where the recovery of the nuclear magnetization, measured by the spin-echo amplitude after the application of an inversion pulse, $M$($t$), recovers as a single-exponential function for $I$ = 3/2, and was fitted by the following function,

\begin{equation}
M(t) = M_0\left[1-c_0\ \mathrm{exp}(-(3t/T_1)^\beta)\right],
\label{SpinLattice}
\end{equation}

\noindent where $M_0$, $c_0$, $t$, and $\beta$ are the equilibum nuclear magnetization, inversion factor ($c_0$ = 2 for complete inversion), the time after the inversion pulse, and the stretching exponent \cite{YamamotoSupplemental}, respectively. We show the temperature dependences of 1/$T_1$ for the $^{69}$Ga-0, $^{69}$Ga-1, and $^{69}$Ga-2 lines in Fig. \ref{TOneRateTempDep}(a). There are three distinct temperature regions where the relaxation processes have different characteristic features: Regions I and II below $T_t$ are in the monoclinic phase, while region III above $T_t$ is in the orthorhombic phase. Before discussing the mechanism in each region, we have to clarify if the origin of the relaxation mechanism is magnetic or quadrupolar. The ratio of $T_1$ for the two isotopes, $^{71}$Ga and $^{69}$Ga, is plotted in Fig. \ref{TOneRateTempDep}(b). As can be seen in the figure, the ratio is 0.6 for all temperatures, which is just equal to ($^{69}\gamma$/$^{71}\gamma$)$^2$, assuring that the relaxation process is governed by magnetic fluctuations, and not by fluctuations of the EFG (a quadrupolar relaxation process would give a ratio of 2.5). In the present case of a purely magnetic relaxation process, the relaxation rate, 1/$T_1$, is generally expressed as \cite{moriya1963effect},

\begin{equation}
\frac{1}{T_1} = \frac{k_BT}{(\gamma_e\hbar)^2}2(\gamma_NA_\perp)^2\sum_{q}f^2(\bm{q})\frac{\mathrm{Im}\chi_{\perp}(\bm{q},\omega_0)}{\omega_0},
\label{TOneRateEq}
\end{equation}

\noindent where $\gamma_e$, Im$\chi_{\perp}$($\bm{q}$,$\omega_0$), and $\omega_0$ are the electronic gyromagnetic ratio, the perpendicular component of the imaginary part of the dynamical susceptibility, $\chi(q,\omega_0)$, with respect to the quantization axis, and the nuclear Larmor frequency, respectively. $f$($\bm{q}$) is the hyperfine form factor. 

For 4$f$ metals such as Ce$_2$Rh$_2$Ga, we have two relaxation channels: one from electron-hole pair excitations across the Fermi level and the other is due to exchange-coupled Ce local spin fluctuations. For the former case, $\chi$($q$,$\omega_0$) is temperature independent and $f$($\bm{q}$) = 1, hence, 1/$T_1$ $\propto$ $T$ (Korringa process) \cite{korringa1950nuclear}, while in the latter case, $\chi$($\bm{q}$,$\omega_0$) may yield a strong $\bm{q}$ and temperature dependence, depending on ferro- or antiferromagnetic correlations. In such a case, we have a temperature-independent relaxation process, 1/$T_1$ $\propto$ $A^2$$J$($J$+1)/3$\omega_{ex}$, where $J\mu_B$ is the Ce localized moment and $\omega_{ex}$ is the exchange frequency, which is estimated from the Weiss constant of the Curie-Weiss (CW) susceptibility \cite{moriya1956nuclear,moriya1956nuclearTwo}. Since the bulk susceptibility of HT-Ce$_2$Rh$_2$Ga shows a CW temperature dependence with a Weiss constant $\theta_p$ $\sim$ -140 K in the orthorhombic phase \cite{nesterenko2020two,YamamotoSupplemental}, the primary relaxation process is considered to stem from AF spin fluctuations. 

$F$($q$), the $q$-summed $f^2$($\bm{q}$), should be calculated from the local geometrical factor at the Ga site \cite{YamamotoSupplemental}. We start to discuss region III, where we see the Korringa behavior 1/$T_1$ $\propto$ $T$. This is not compatible with the localized picture of exchange-coupled 4$f$ Ce$^{3+}$ moments. We calculate that the hyperfine form factor $F$($\pi$, $\pi$, $\pi$) is approximately 1/64 of $F$(0, 0, 0), by assuming an isotropic hyperfine coupling constant between the Ga and Ce ions in the orthorhombic phase, taking into account the Ce ions up to the third-nearest neighbor of the Ga ion \cite{YamamotoSupplemental}. Correspondingly, the contribution of AF fluctuations to 1/$T_1$ is strongly suppressed due to the form factor in region III, and the Korringa process exceeds the contribution by AF fluctuations, which results in the found linear temperature dependence of 1/$T_1$. 

Because of the lowering of the local symmetry at the Ga sites below $T_t$ (monoclinic phase), one can expect that the AF fluctuations are not screened in region II, which results in the temperature-independent relaxation process via fluctuations of the local Ce moment contributions at the Ga-1 and Ga-2 sites. In region I, 1/$T_1$ exhibits again a Korringa behavior below 0.8 K for Ga-1 and below 2.0 K for Ga-2. This crossover of the relaxation process firmly indicates that the Ce 4$f$ state is strongly hybridized with the conduction electrons, forming a coherent heavy-electron state, i.e., a dense Kondo coherent state. This formation of a Kondo coherent state has also been reported from measurements of temperature-dependent 1/$T_1$ for other Ce-based heavy fermions \cite{kawasaki1998si}. It is worth to point out that the coherence temperature is different for Ga-1 ($\sim$ 0.8 K) and Ga-2 ($\sim$ 2.0 K). This presumably results from different Kondo screening, i.e., different Kondo temperatures for the two inequivalent Ce ions. Specific-heat results obtained on HT-Ce$_2$Rh$_2$Ga polycrystals also indicate the formation of a heavy-fermion state with a Sommerfeld coefficient $\gamma$ = 1 J/mol$_\mathrm{f.u.}$ K$^2$ \mbox{\cite{YamamotoSupplemental}}.

\textit{Conclusion.} --- We investigated the microscopic nature of the 130-K phase transition and the formation of the unusual ground state of HT-Ce$_2$Rh$_2$Ga using NQR. The spectral variation with temperature across  $T_t$ $\sim$ 130 K provides evidence for the absence of magnetic or charge order. The temperature dependence of 1/$T_1$ is governed by the antiferromagnetically coupled Ce-4$f$ local spin fluctuations, but both above $T_t$, and below 0.8 K for the Ga-1 site and below 2.0 K for the Ga-2 site, 1/$T_1$(T) behaves Korringa-like. Here, we argue that the former is due to the screening of the AF fluctuations by the hyperfine form factor, and the latter is due to the formation of a Kondo coherent state. Particularly, the emergence of a two-ion Kondo state with inequivalent Ce ions being screened at different temperatures is the most important outcome of the present study. It can be assumed that the Fermi surface is very anisotropic in Ce$_2$Rh$_2$Ga, which makes future studies of single crystals very interesting. A spatially anisotropic conductivity limits the RKKY interaction in real space, and, thus, affects the competition with the isotropic Kondo interaction. This creates an anisotropic Kondo system with multi-ion Kondo physics, which is certainly an exciting topic for future studies.

\section*{Acknowledgments}
We thank A. Tursina for fruitful discussions and for leading our understanding of the crystal structure of the title compound. U. Nitsche is acknowledged for technical support. We acknowledge the support of the HLD at HZDR, member of the European Magnetic Field Laboratory (EMFL), the Deutsche Forschungsgemeinschaft (DFG) through SFB 1143, and the W\"{u}rzburg-Dresden Cluster of Excellence on Complexity and Topology in Quantum Matter--$ct.qmat$ (EXC 2147, Project ID 390858490). A.M.S. thanks the NRF (93549) and the URC/FRC of UJ for financial assistance.

\end{document}


\enlargethispage{2\baselineskip}
\title{}

\begin{center}
\noindent\textbf{{\Large Supplemental Material}}
\end{center}

\noindent\textbf{{\Large Structural 130-K Phase Transition and Emergence of a Two-Ion Kondo State in HT-Ce$_2$Rh$_2$Ga Explored by $^{69,71}$Ga Nuclear Quadrupole Resonance}}

\author{Sh. Yamamoto}
\email{s.yamamoto@hzdr.de}
\affiliation{Hochfeld-Magnetlabor Dresden (HLD-EMFL) and W\"{u}rzburg-Dresden Cluster of Excellence ct.qmat, Helmholtz-Zentrum Dresden-Rossendorf, 01328 Dresden, Germany}
\affiliation{Max Planck Institute for Chemical Physics of Solid, D-01187 Dresden, Germany}
\author{T. Fujii}
\affiliation{Max Planck Institute for Chemical Physics of Solid, D-01187 Dresden, Germany}
\author{S. Luther}
\affiliation{Hochfeld-Magnetlabor Dresden (HLD-EMFL) and W\"{u}rzburg-Dresden Cluster of Excellence ct.qmat, Helmholtz-Zentrum Dresden-Rossendorf, 01328 Dresden, Germany}
\affiliation{Institut f$\ddot{u}$r Festk$\ddot{o}$rper- und Materialphysik, TU Dresden, 01062 Dresden, Germany}
\author{H. Yasuoka}
\affiliation{Max Planck Institute for Chemical Physics of Solid, D-01187 Dresden, Germany}
\author{H. Sakai}
\affiliation{Advanced Science Research Center, Japan Atomic Energy Agency, Tokai, Ibaraki 319-1195, Japan}
\author{F. B\"{a}rtl}
\affiliation{Hochfeld-Magnetlabor Dresden (HLD-EMFL) and W\"{u}rzburg-Dresden Cluster of Excellence ct.qmat, Helmholtz-Zentrum Dresden-Rossendorf, 01328 Dresden, Germany}
\affiliation{Institut f$\ddot{u}$r Festk$\ddot{o}$rper- und Materialphysik, TU Dresden, 01062 Dresden, Germany}
\author{K. M. Ranjith}
\affiliation{Max Planck Institute for Chemical Physics of Solid, D-01187 Dresden, Germany}
\author{H. Rosner}
\affiliation{Max Planck Institute for Chemical Physics of Solid, D-01187 Dresden, Germany}
\author{J. Wosnitza}
\affiliation{Hochfeld-Magnetlabor Dresden (HLD-EMFL) and W\"{u}rzburg-Dresden Cluster of Excellence ct.qmat, Helmholtz-Zentrum Dresden-Rossendorf, 01328 Dresden, Germany}
\affiliation{Institut f$\ddot{u}$r Festk$\ddot{o}$rper- und Materialphysik, TU Dresden, 01062 Dresden, Germany}
\author{A. M. Strydom}
\affiliation{Max Planck Institute for Chemical Physics of Solid, D-01187 Dresden, Germany}
\affiliation{Highly Correlated Matter Research Group, Physics Department, University of Johannesburg, PO Box 524, Auckland Park 2006, South Africa}
\author{H. K\"{u}hne}
\affiliation{Hochfeld-Magnetlabor Dresden (HLD-EMFL) and W\"{u}rzburg-Dresden Cluster of Excellence ct.qmat, Helmholtz-Zentrum Dresden-Rossendorf, 01328 Dresden, Germany}
\author{M. Baenitz}
\affiliation{Max Planck Institute for Chemical Physics of Solid, D-01187 Dresden, Germany}

\date{\today}
\maketitle

We present here:

\begin{enumerate}
\item Crystal structure of HT-Ce$_2$Rh$_2$Ga
\item Experimental details of the NQR measurements
\item Calculation of the form factor in the orthorhombic phase of HT-Ce$_2$Rh$_2$Ga
\item Macroscopic data [susceptibility ($\chi$ vs $T$), magnetization ($M$ vs $H$), and specific heat ($C$ vs $T$ and C$_m$/$T$ vs $T$)] of polycrystalline HT-Ce$_2$Rh$_2$Ga
\end{enumerate}

\newpage
\noindent{\large SUPPLEMENTARY NOTE 1:}

\vspace{3pt}

\noindent{\large Crystal structure of HT-Ce$_2$Rh$_2$Ga}

\vspace{14pt}

\begin{figure}[H]
	\begin{center}
		\includegraphics[width=13.5cm]{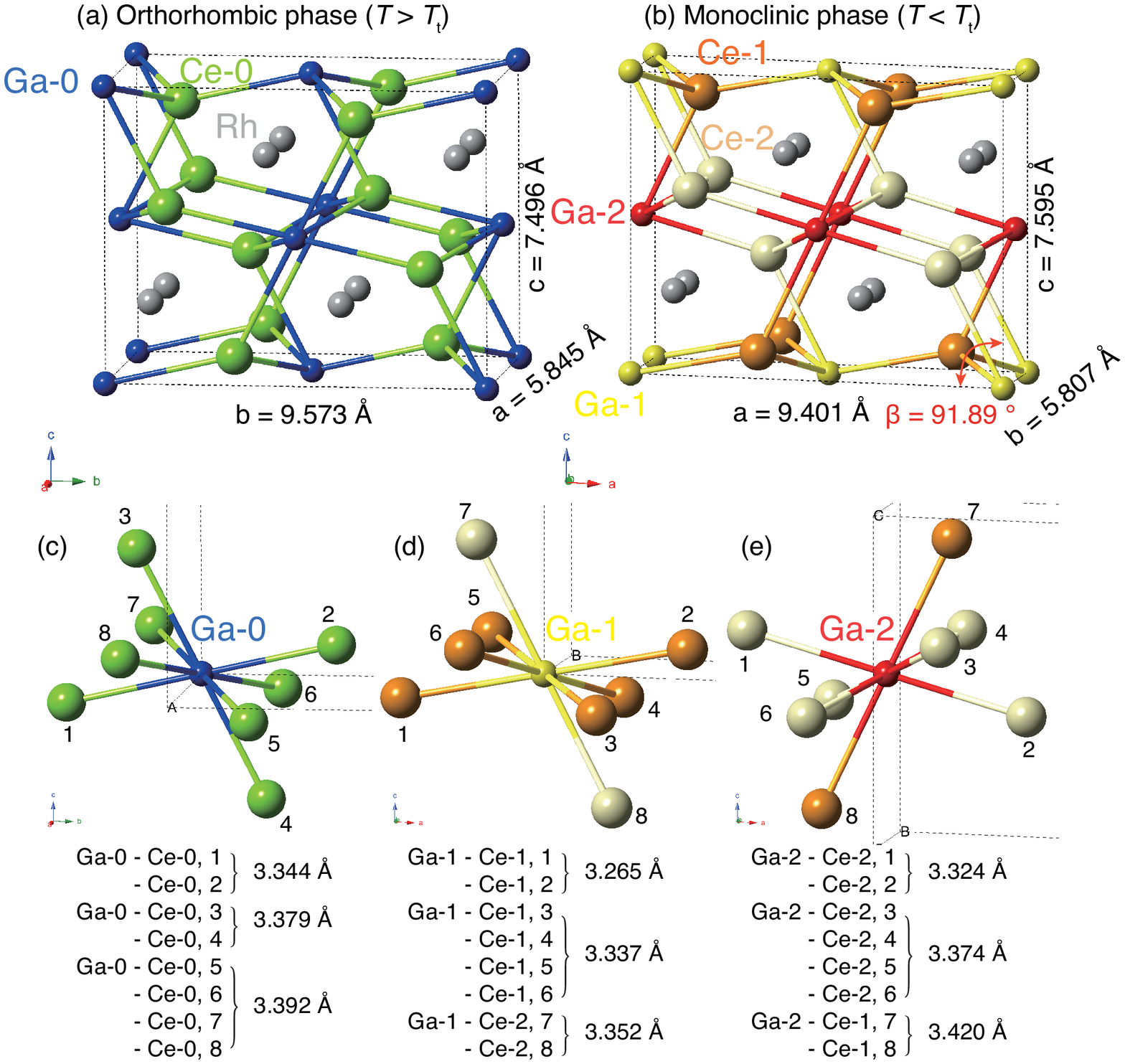}
	\end{center}
	\caption{Crystal structure of HT-Ce$_2$Rh$_2$Ga. (a) Orthorhombic phase above $T_t$ $\sim$ 128 K with La$_2$Ni$_3$-type (space group $Cmce$). (b) Monoclinic phase (space group $C2/m$) below $T_t$. Local coordination environment of 1st, 2nd, and 3rd nearest neighboring Ce atoms when viewed from (c) a Ga-0 (4$a$) atom in the orthorhombic $Cmce$, and (d) from a Ga-1 (2$a$) and (e) Ga-2 (2$d$) atom in the monoclinic $C2/m$ phases. The bond lengths are also listed in (c), (d), and (e).}
	\label{Fig:CrystalStructure}
\end{figure}

HT-Ce$_2$Rh$_2$Ga crystallizes in the orthorhombic structure of La$_2$Ni$_3$ type (space group $Cmce$) above $T_t$ $\sim$ 130 K \cite{nesterenko2020two}. Below $T_t$, the crystal structure changes to monoclinic (space group $C2/m$) being accompanied by non-merohedral twinning \cite{dudka2022multi}. Above $T_t$, there are single crystallographic sites for the Ce and Ga ions, while below $T_t$, there are two inequivalent Ce and Ga sites (Fig. \ref{Fig:CrystalStructure}). Bond lengths around Ga-0 in the orthorhombic $Cmce$, and Ga-1 (2$a$) and Ga-2 (2$d$) in the monoclinic $C2/m$ phases are shown in Figs. \mbox{\ref{Fig:CrystalStructure}}(c), \mbox{\ref{Fig:CrystalStructure}}(d), and \mbox{\ref{Fig:CrystalStructure}}(e). Passing through the crystallographic phase transition from the high-temperature orthorhombic to low-temperature monoclinic phase at $\sim$ 130 K, the average Ga-Ce bond length is reduced by 1.60 \% and 0.12 \% for the Ga-1 and  Ga-2 sites, respectively.


\newpage
\noindent{\large SUPPLEMENTARY NOTE 2:}

\vspace{3pt}

\noindent{\large Experimental details of the NQR measurements}

\vspace{14pt}

The $^{69,71}$Ga NQR experiments were carried out with a high-quality powder sample prepared from the arc-melted single crystal. Using a 200 $\mu$m sieve, the particle size were selected to be less than  200 $\mu$m, and the obtained powder was mixed with paraffin. The NQR spectra and the nuclear magnetic relaxation times were measured using a standard pulsed (spin-echo) NMR apparatus (TecMag-Apollo).

The NQR spectra were taken by the frequency-sweep method under zero applied magnetic field. In order to avoid any artificial broadening, the real part of the spin-echo signal was integrated after proper phase adjustments at each frequency. The typical excitation-pulse width, two-pulse separation and repetition ($t_r$) time were 6 $\mu$sec, 30 $\mu$sec, and 100 msec, respectively. After several nutation experiments, the 6 $\mu$sec - 12 $\mu$sec pulse sequence was found to be the best for forming the spin-echo. From the typical 6 $\mu$sec exciting pulse width using a Cu coil (13-mm length and 2-mm radius), we can estimate that  the radio-frequency (rf) field acting on the sample is of the order of 40 Oe. We used the temperature steps of 5 K and 10 K around $T_t$ in our NQR spectral measurements.

As to a possible heating of the sample during the experiments, we calculated the skin-depth for a 30 MHz rf-field from the resistivity of Ce$_2$Rh$_2$Ga in the monoclinic phase ($\sim$ 300 $\mu\Omega$ cm) to be about 160 $\mu$m which is just about the maximum particle size of the sample. Therefore, the skin-depth effect due to eddy currents induced by rf excitation during the NQR measurements is considered to be negligible. We took special care not to heat the sample during the measurements, particularly in the $^3$He temperature range. At selected temperatures, we measured the $t_r$ dependence of the spin-echo (SE) intensity. Above a certain value $t^T_r$, the SE intensity becomes independent of $t_r$, where the increase of the sample temperature due to radio-frequency excitation is negligible. We used $t_r$ which is at least twice as large as $t^T_r$.

\begin{figure}[H]
	\begin{center}
		\includegraphics[width=17cm]{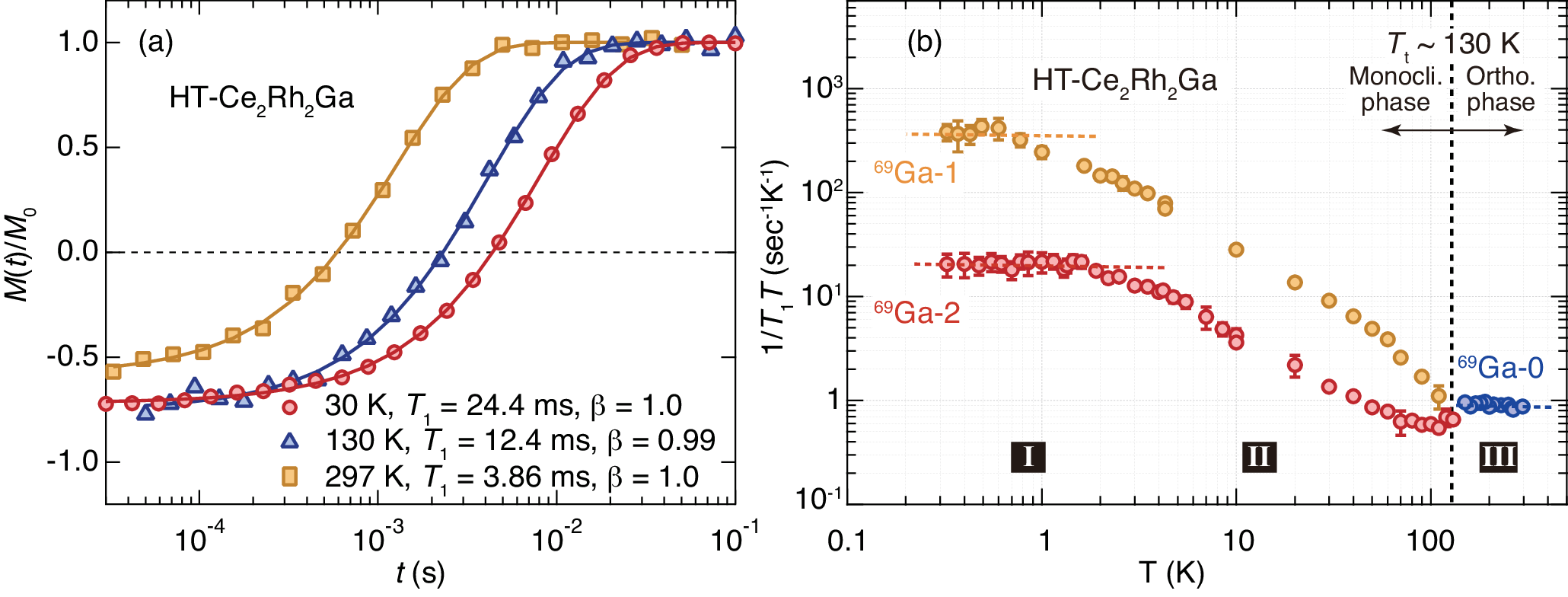}
	\end{center}
	\caption{(a) Recovery of $^{69}$Ga nuclear magnetization $M$($t$) divided by the saturation value $M_0$ as a function of the delay $t$ between the inversion and the first spin-echo pulses. The yellow, blue, and red symbols are data at 30 K, 130 K, and 297 K, which are below, close to, and above $T_t$, respectively. The solid lines are fits of recovery curves using Eq. (4). The stretching exponent $\beta$ is $\sim$ 1 at each temperature. (b) Temperature dependence of 1/$T_1T$, for the $^{69}$Ga-0 (blue), $^{69}$Ga-1 (yellow) and $^{69}$Ga-2 (red) lines.}
	\label{Fig:ToneResults}
\end{figure}

The temperature dependence of the nuclear magnetic relaxation rate, 1/$T_1$, was measured by the inversion-recovery method, where the evolution of the spin-echo amplitude after an inversion pulse was recorded as a function of the time interval between the inversion pulse and the first spin-echo pulse. The typical examples of the recovery curves are shown in Fig. \ref{Fig:ToneResults}(a) at 30 K (monoclinic phase), 130 K (close to the phase transition), and 297 K (orthorhombic phase). For $I$ = 3/2, the recovery curve is described by Eq. (4) in the main text. Fig. \ref{Fig:ToneResults}(a) shows that all recovery curves are well described with a stretching exponent of $\beta$ = 1.0, meaning the $T_1$ values are determined by a single relaxation process and no inhomogeneity is present at all temperatures. The error bars in Fig. 3(a) in the main text were extracted from the standard deviation of the fits to the recovery curves using Eq. (4). Figure \ref{Fig:ToneResults}(b) shows the temperature dependence of 1/$T_1T$. Three distinct temperature regions are seen.


\newpage
\noindent{\large SUPPLEMENTARY NOTE 3:}

\vspace{3pt}

\noindent{\large Calculation of the form factor of hyperfine coupling at the Ga sites of HT-Ce$_2$Rh$_2$Ga above $T_t$}

\vspace{14pt}

\begin{figure}[H]
	\begin{center}
		\includegraphics[width=10cm]{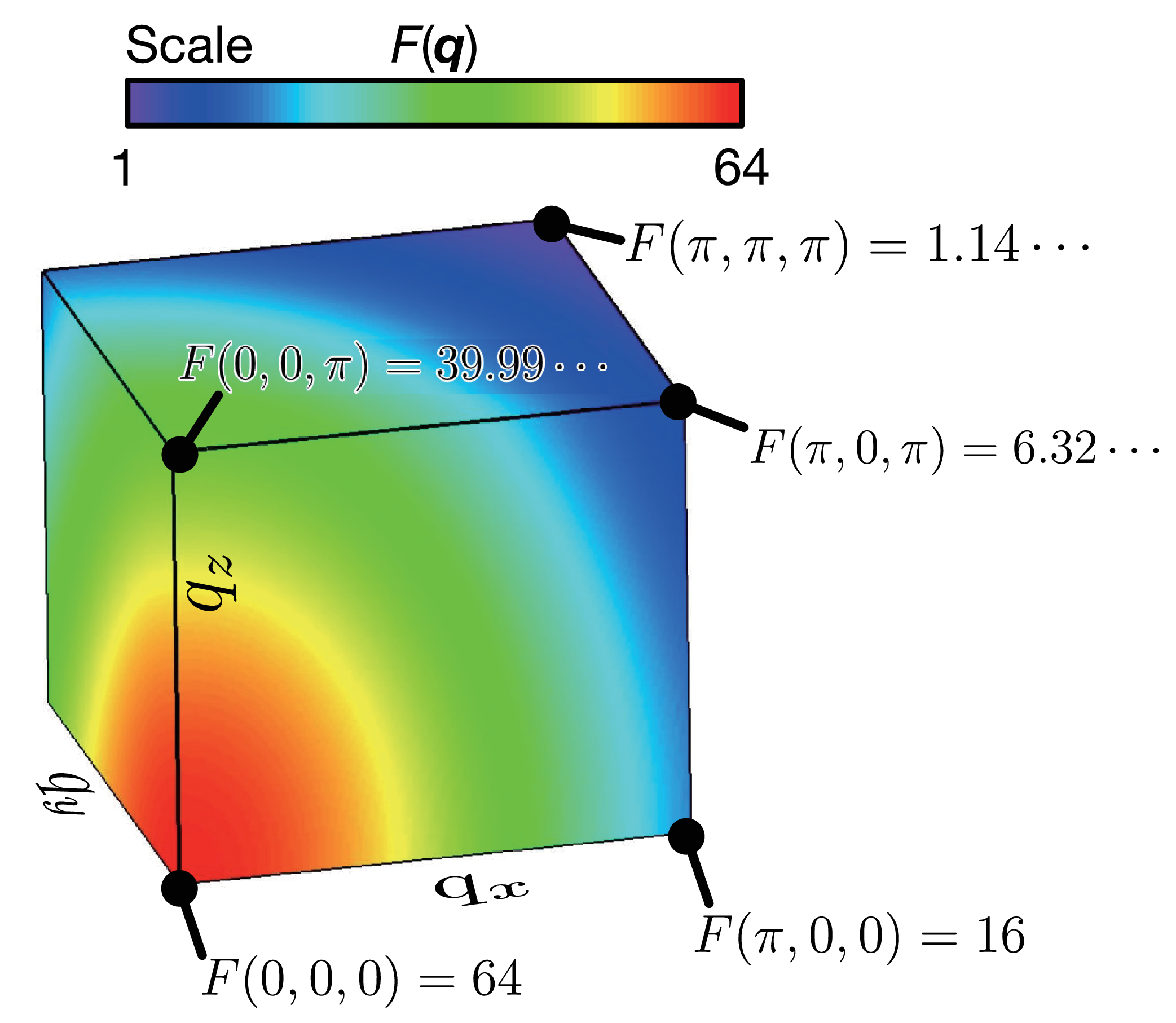}
	\end{center}
	\caption{Color scale plot of the Ga form factor in the unit of $A_0^2$, where $A_0$ is the transferred hyperfine coupling constant. The $A_0$ is assumed to be uniform from the eight Ce-0 atoms [see Fig. \ref{Fig:CrystalStructure}(c)].}
	\label{Fig:FormFactor}
\end{figure}

In Eq. (5) of the main text, the $\bm{q}$-dependent hyperfine term $A_{\perp}^2\Sigma_{\bm{q}}f^2(\bm{q})$ can often be approximated to $B_{0}^2F(\bm{q})$, where $B_{0}$ is the averaged hyperfine coupling constant from the neighboring magnetic atoms and $F(\bm{q})$ is the $\bm{q}$-summed $f^2$($\bm{q}$). $F(\bm{q})$ can be estimated from the local geometry as follows. For many intermetallic compounds, the hyperfine field is transferred by a RKKY-type mechanism with the local spins via conduction electrons. In such a case, the transferred hyperfine coupling constant would be inversely proportional to the cube of the distance ($\propto 1/r_{i}^3$) from the local spin. In HT-Ce$_2$Rh$_2$Ga, the Ga sites are located at the center of eight Ce atoms, as illustrated in Fig. \ref{Fig:CrystalStructure}(c). Assuming the same magnitude of $B_{0}$ from the surrounding Ce atoms, $F(\bm{q})$ can be numerically calculated from $\large\{\Sigma_{i}\exp(i\bm{q}\cdot\bm{r}_i)\large\}^2$, where $\bm{r}_i$ is the local coordinate of each Ce atom with regard to the Ga sites.  The calculated result is visualized in Fig. \ref{Fig:FormFactor}. $F(\bm{q})$ becomes much smaller at the AFM positions of $q_x=\pi$, $q_y=\pi$, and $q_z=\pi$ compared to $F(\bm{q}=0)$, which results in the suppression of the contribution of AF fluctuations to 1/$T_1$ in region III.

\newpage
\noindent{\large SUPPLEMENTARY NOTE 4:}

\vspace{3pt}

\noindent{\large Macroscopic data of polycrystalline HT-Ce$_2$Rh$_2$Ga}

\vspace{14pt}

\begin{figure}[H]
	\begin{center}
		\includegraphics[width=16cm]{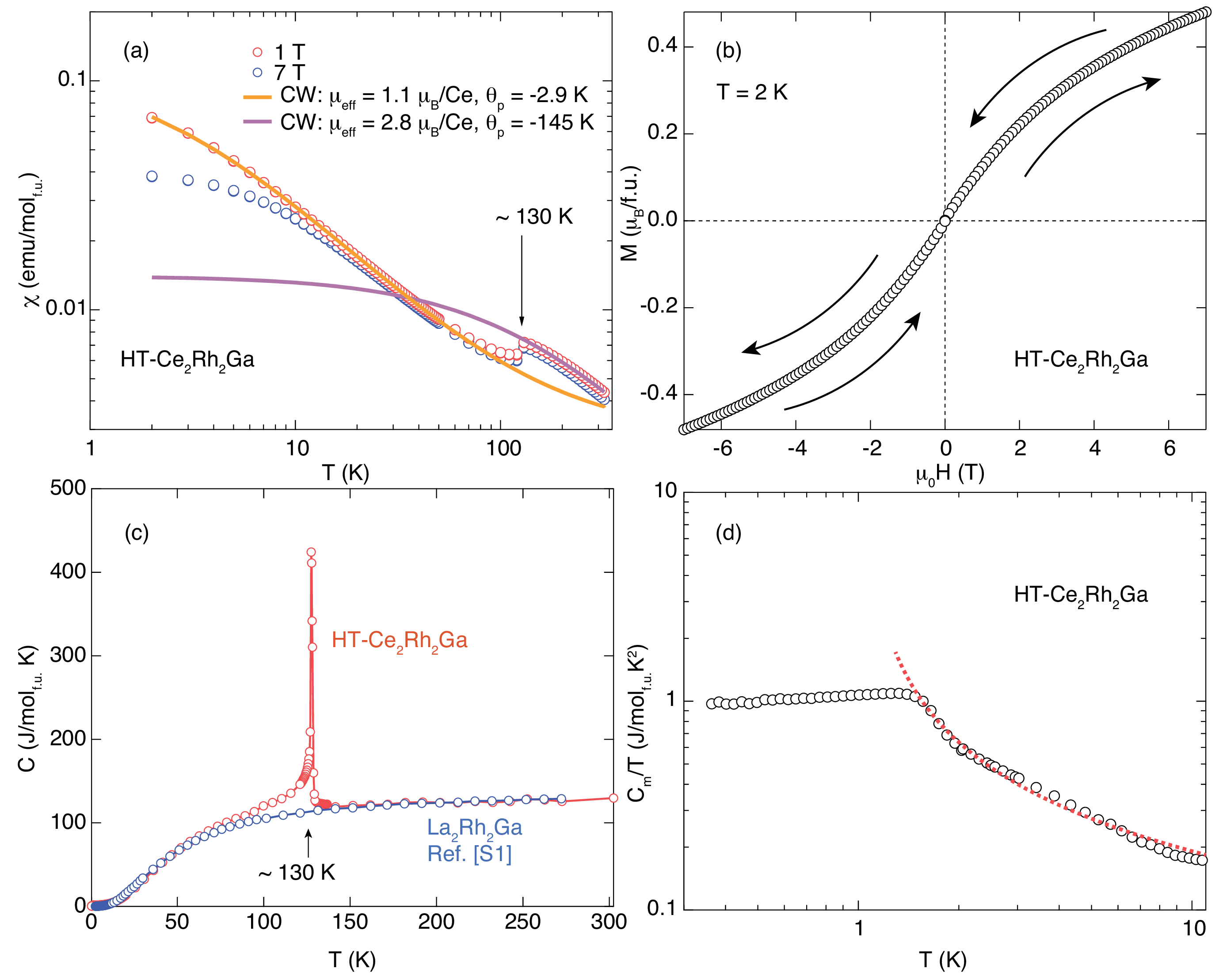}
	\end{center}
	\caption{(a) Susceptibility of polycrystalline HT-Ce$_2$Rh$_2$Ga as a function of temperature in magnetic fields of 1 and 7 T. The orange (purple) solid line represents a Curie-Weiss fit to the measured data below 50 K (above 150 K). The structural phase transition at $\sim$ 130 K is clearly visible. (b) Magnetization of polycrystalline HT-Ce$_2$Rh$_2$Ga as a function of cycled magnetic fields between -7 T and +7 T at 2 K. No hysteresis associated with long-range order is visible. (c) Temperature dependence of the specific heat, $C$, for HT-Ce$_2$Rh$_2$Ga and its nonmagnetic relative La$_2$Rh$_2$Ga \cite{nesterenko2020two}. The low-temperature magnetic specific heat, $C_m$, of Ce$_2$Rh$_2$Ga is determined by subtracting the contribution of La$_2$Rh$_2$Ga. (d) Extracted magnetic contribution $C_m$/$T$ plotted against $T$ in a log-log scale. The dashed curve is an empirical fit of the data to $c$/log($T$) with $c$ = 0.19. $C_m$/$T$ is temperature independent below 1.4 K with the electronic specific heat, $\gamma$ = 1 J/mol$_\mathrm{f.u.}$K$^2$.}
	\label{Fig:MacroData}
\end{figure}

Among various macroscopic measurements on the sample used in the present NQR study, we show the temperature dependence of the static magnetic susceptibility [Fig. \ref{Fig:MacroData}(a)], a lack of magnetic hysteresis at 2 K [Fig. \ref{Fig:MacroData}(b)], and specific-heat measurements [Figs. \ref{Fig:MacroData}(c) and \ref{Fig:MacroData}(d)]. The magnitude of the Weiss constant $\theta_p$ obtained from the Curie-Weiss fit drastically changes across the structural phase transition. In addition, the Ce valence increases by $\sim$ 0.7 \% below $T_t$ \cite{sato2022valence}. These two experimental results evidence the reduction of the antiferromagnetic exchange interaction between the Ce ions below $T_t$. It should be emphasized here that the nearly temperature-independent magnetic specific heat, $C_m$/$T$,  below 1.4 K provides strong evidence of the presence of a heavy-Fermi liquid state with $\gamma$ = 1 J/mol$_\mathrm{f.u.}$K$^2$ [Fig. \ref{Fig:MacroData}(d)].

\vspace{14pt}

